\documentclass[aps,prb,twocolumn,letterpaper]{revtex4}

\usepackage{amsmath,amssymb,amsfonts,amsthm}
\usepackage{graphicx,bm}

\begin{document}

\title{Origin of ferroelectricity in the multiferroic barium fluorides Ba$\bm
  M$F$_4$}

\date{\today}

\author{Claude Ederer}
\affiliation{Materials Department, University of California, Santa
  Barbara, CA 93106, U.S.A.}
\email{ederer@mrl.ucsb.edu}
%\homepage{http://www.mrl.ucsb.edu/~ederer}
\author{Nicola A.~Spaldin}
\affiliation{Materials Department, University of California, Santa
  Barbara, CA 93106, U.S.A.}
 
\begin{abstract}
  We present a first principles study of the series of multiferroic barium
  fluorides with the composition Ba$M$F$_4$, where $M$ is Mn, Fe, Co, or Ni.
  We discuss trends in the structural, electronic, and magnetic properties,
  and we show that the ferroelectricity in these systems results from the
  ``freezing in'' of a single unstable polar phonon mode. In contrast to the
  case of the standard perovskite ferroelectrics, this structural distortion
  is not accompanied by charge transfer between cations and anions. Thus, the
  ferroelectric instability in the multiferroic barium fluorides arises solely
  due to size effects and the special geometrical constraints of the
  underlying crystal structure.
\end{abstract}

\pacs{}

\maketitle

\section{Introduction}

Magnetoelectric multiferroics are currently attracting much
attention.\cite{Fiebig:2005, Spaldin/Fiebig:2005} These materials, which
exhibit magnetic and dielectric order in the same phase, can give rise to
interesting coupling effects between the two ferroic order parameters, with
great potential for technological applications. Currently, the main research
effort in multiferroics is directed towards Bi-containing perovskites such as
BiFeO$_3$ or BiMnO$_3$,\cite{Wang_et_al:2003, Neaton_et_al:2005,
  Hill/Rabe:1999} and towards both hexagonal and orthorhombic rare-earth
manganites such as HoMnO$_3$ and TbMnO$_3$.\cite{Lottermoser_et_al:2004,
  Kimura_et_al_Nature:2003} In the present article we revisit another class of
magnetic ferroelectrics: the barium fluorides Ba$M$F$_4$, where $M$ is Mn, Fe,
Co, or Ni. This series of compounds attracted considerable interest from the
late 1960s until the early 1980s (see Ref.~\onlinecite{Scott:1979} and
references therein) but so far has not received much attention during the
recent revival of multiferroic materials. We hope that our work will inspire
further research in this interesting class of multiferroic materials, with the
goal of exploring a broader range of non-oxide-based materials as candidates
for magnetoelectric device applications.

The barium fluorides Ba$M$F$_4$, $M$ = Mn, Fe, Co, Ni, Zn, and Mg, form an
isostructural family of compounds with the polar space group $Cmc2_1$
(Refs.~\onlinecite{Schnering/Bleckmann:1968,
  Eibschuetz_et_al:1969}).\footnote{In many papers this space group is denoted
  as $A2_1am$, referring to an alternative setting for the base-centered
  orthorhombic lattice vectors. Here, we use the notation of
  Ref.~\onlinecite{Bradley/Cracknell:Book}.} The corresponding crystal
structure is shown in Fig.~\ref{fig:struc}a. In this base-centered
orthorhombic structure the transition metal cations are octahedrally
surrounded by fluorine anions. Four of the six corners of the fluorine
octahedra are shared with adjacent octahedra to form puckered sheets
perpendicular to the orthorhombic $b$ axis. These sheets of octahedra are
separated by similar sheets of Ba cations. The structure is polar along the
$c$ direction, and ferroelectric switching has been demonstrated for $M$ = Co,
Ni, Zn, and Mg, but not for $M$ = Mn and Fe.\cite{Eibschuetz_et_al:1969} The
dielectric constants $\epsilon_c$ show an increase with temperature that is
characteristic for a ferroelectric phase transition, but all crystals melt
before a transition into the paraelectric phase occurs. The melting
temperatures range between 720--965\,$^\circ$C; the Curie-Weiss temperatures
can be extrapolated from the temperature dependence of the dielectric
constants and range between 810--1320\,$^\circ$C.\cite{DiDomenico_et_al:1969}
In BaMnF$_4$ an additional structural phase transition occurs at
$\sim$255~K.\cite{Spencer/Guggenheim/Kominiak:1970} The resulting
low-temperature structure is incommensurate along the $c$
axis.\cite{Shapiro_et_al:1976} To facilitate a systematic comparison between
the different Ba$M$F$_4$ systems we do not consider this low temperature
structure of BaMnF$_4$ in the present study and instead use perfect $Cmc2_1$
symmetry for all systems.

In addition to the polar distortion, the systems with $M$ = Mn, Fe, Co, and Ni
exhibit antiferromagnetic ordering below $T_\text{N} \approx$ 50 -
120\,$^\circ$C.\cite{Eibschuetz/Guggenheim:1968} These materials therefore
exhibit multiferroic behavior, which is the motivation for the present
investigation.

Here we present a comprehensive computational study of the structural,
electronic, magnetic, and ferroelectric properties of Ba$M$F$_4$, with $M$ =
Mn, Fe, Co, and Ni, using first principles electronic structure calculations.
The goal is to elucidate the origin of ferroelectricity in these materials,
and to understand trends, such as why the systems with $M$ = Mn and Fe do not
exhibit ferroelectric switching. The origin of ferroelectricity is of
fundamental interest, since Ba$M$F$_4$ does not contain any ions, which are
usually considered to be ``ferroelectrically active'', such as empty $d$ shell
cations or lone pair-active cations (e.g. Bi$^{3+}$,
Pb$^{2+}$).\cite{Hill:2000,Seshadri/Hill:2001} We show that the
ferroelectricity in these systems is due to the softening of a single polar
phonon mode, which involves both rotational motions of the fluorine octahedra
and polar displacements of the Ba cations. The instability is caused solely by
size effects and geometrical constraints; no charge transfer between anions
and cations occurs as a result of the structural distortion. The Ba$M$F$_4$
multiferroics therefore represent an example of proper ``geometric
ferroelectricity'', a mechanism that has been proposed as a possible way to
incorporate both magnetism and ferroelectricity in the same
system.\cite{vanAken_et_al:2004, Ederer/Spaldin:2004} Furthermore, our work
represents the first \emph{ab initio} study of a non-oxide multiferroic
system, and we provide evidence that the LSDA+$U$ method results in a good
description of the electronic structure of these materials.

This paper is organized as follows. We first describe the methods we use in
our calculations, together with some technical details. We then present the
results of our calculations for the structural, electronic, and magnetic
properties of all systems.  Finally, we focus on the ferroelectric properties
and analyze the mechanism underlying the polar structural distortions in these
systems. We end with a discussion and summary.

\section{Computational Method}
\label{sec:method}

All calculations in this work are performed using the projector augmented-wave
method,\cite{Bloechl:1994} implemented in the {\sc Vienna Ab-Initio Simulation
  Package} (VASP).\cite{Kresse/Furthmueller_PRB:1996, Kresse/Joubert:1999} For
the treatment of exchange and correlation we use both the local spin-density
approximation (LSDA) \cite{Jones/Gunnarsson:1989} and the LSDA+$U$ method in
the formulation of Dudarev {\it et al.},\cite{Dudarev_et_al:1998} which is
equivalent to the standard form of Anisimov {\it et al.},
\cite{Anisimov/Aryatesiawan/Liechtenstein:1997} with the intra-atomic exchange
parameter $J$ set to zero (see Ref.~\onlinecite{Ederer/Spaldin_2:2005}). For
the Hubbard parameter $U_\text{eff} = U - J$ of the transition metal $d$
states we use a typical value of 4\,eV, and we test the sensitivity of our
results with respect to the precise value of $U_\text{eff}$ where necessary
(see also section~\ref{sec:ldau}).

To obtain structural parameters we relax all ions until the Hellman-Feynman
forces are less than 0.01\,eV/\AA, and we adjust the lattice vectors such that
all components of the stress tensor are smaller than 1\,kbar. During the
relaxations, except where otherwise noted, we impose the experimentally
observed antiferromagnetic order (see Sec.~\ref{sec:mag}), which doubles the
size of the unit cell and reduces the space group symmetry to monoclinic.
However, no monoclinic distortion has been found experimentally, and in our
calculations the corresponding effect, if present, is too small to be resolved
unambiguously. We therefore neglect a possible monoclinic distortion and
restrict the lattice vectors to the orthorhombic symmetry. For test purposes,
we also perform relaxations for different magnetic structures, e.g.
ferromagnetic ordering.

To obtain local densities of states we define spheres around the ions with
radius 1.0\,\AA\ for the $M$ cations and 0.9\,\AA\ for the fluorine anions. We
use a 4$\times$4$\times$3 Monkhorst-Pack k-point mesh (divisions with respect
to the base-centered orthorhombic lattice vectors of the nonmagnetic unit
cell) and a Gaussian smearing of 0.1\,eV for Brillouin zone integrations. The
plane-wave energy cutoff is set to 550\,eV for relaxations and to 450\,eV for
all other calculations. To calculate the spontaneous polarization we use the
Berry-phase approach \cite{King-Smith/Vanderbilt:1993,
  Vanderbilt/King-Smith:1993, Resta:1994} and integrate over 8 homogeneously
distributed k-point strings parallel to the reciprocal $c$ direction, each
string containing 8 k-points. For the calculation of the force-constant matrix
we displace all the ions by 0.005\,\AA, corresponding to symmetry-adapted
modes. To exclude eventual nonlinearities in the forces we repeat the
calculation with displacements of 0.01\,\AA, which leads to identical values
for the force constant matrix.

\section{Results and Discussion}

\subsection{Structural properties}
\label{sec:struc}

\begingroup\squeezetable
\begin{table*}
\caption{Structural parameters for $Cmc2_1$ Ba$M$F$_4$, $M$=Mn, Fe,
  Co, Ni, obtained using the LSDA and the LSDA+$U$ method with
  $U_\text{eff}$=4~eV together with experimental data. $a$, $b$, $c$ 
  are the usual orthorhombic lattice constants, $V$ is the
  corresponding volume, $d$ represents the atomic displacements
  compared to the $Cmcm$ centrosymmetric reference structure. All
  atomic positions correspond to Wyckoff-positions 4$a$: (0,$y$,$z$).}
\label{tab:struc}
\begin{ruledtabular}
\begin{tabular}{lc|ccc|ccc|ccc|ccc}
 & & \multicolumn{3}{c|}{Mn} & \multicolumn{3}{c|}{Fe} & \multicolumn{3}{c|}{Co} & \multicolumn{3}{c}{Ni} \\
 & & LSDA & $U_\text{eff}$=4~eV & Expt. [\onlinecite{Keve/Abrahams/Bernstein:1969}] & LSDA &
$U_\text{eff}$=4~eV & Expt. [\onlinecite{Averdunk/Hoppe:1988}] & LSDA & $U_\text{eff}$=4~eV &
Expt. [\onlinecite{Keve/Abrahams/Bernstein:1970}] & LSDA & $U_\text{eff}$=4~eV
 & Expt. [\onlinecite{Welsch_et_al:1999}] \\
\hline
& $a$ [\AA] & 4.18 & 4.18 & 4.22 & 4.20 & 4.21 & 4.24 & 4.15 & 4.10 & 4.21 & 4.08 & 4.08 & 4.14 \\
& $b$ [\AA] & 14.58 & 14.70 & 15.10 & 14.30 & 14.52 & 14.86 & 14.05 & 14.08 & 14.63 & 13.85 & 13.95 & 14.43 \\
& $c$ [\AA] & 5.81 & 5.85 & 5.98 & 5.63 & 5.59 & 5.83 & 5.65 & 5.78 & 5.85 & 5.65 & 5.67 & 5.78 \\
& $V$ [\AA$^3$] & 354.4 & 359.5 & 381.4 & 338.0 & 341.7 & 367.2 & 328.9 & 333.4 & 360.3 & 319.4 & 323.0 & 345.1 \\
\hline
Ba   & $y$ & 0.152 & 0.154 & 0.156 & 0.145 & 0.148 & 0.151 & 0.142 & 0.143 & 0.148 & 0.139 & 0.141 & 0.146 \\
     & $z$ & -0.043 & -0.047 & -0.047 & -0.035 & -0.034 & -0.041 & -0.032 & -0.040 & -0.039 & -0.026 & -0.031 & -0.036 \\
%\hline
Ni   & $y$ & 0.414 & 0.414 & 0.416 & 0.410 & 0.411 & 0.414 & 0.409 & 0.410 & 0.413 & 0.408 & 0.409 & 0.412 \\
     & $z$ & 0.000 & 0.000 & 0.000 & 0.000 & 0.000 & 0.000 & 0.000 & 0.000 & 0.000 & 0.000 & 0.000 & 0.000 \\
%\hline
F(1) & $y$ & -0.464 & -0.462 & -0.465 & -0.471 & -0.468 & -0.469 & -0.477 & -0.474 & -0.472 & -0.483 & -0.479 & -0.475 \\
     & $z$ & -0.158 & -0.157 & -0.163 & -0.171 & -0.157 & -0.166 & -0.191 & -0.191 & -0.179 & -0.210 & -0.200 & -0.187 \\
%\hline
F(2) & $y$ & 0.298 & 0.298 & 0.298 & 0.299 & 0.300 & 0.301 & 0.300 & 0.299 & 0.302 & 0.302 & 0.301 & 0.303 \\
     & $z$ & 0.205 & 0.202 & 0.196 & 0.212 & 0.208 & 0.198 & 0.213 & 0.204 & 0.198 & 0.217 & 0.211 & 0.202 \\
%\hline
F(3) & $y$ & -0.327 & -0.329 & -0.336 & -0.324 & -0.323 & -0.331 & -0.325 & -0.329 & -0.334 & -0.322 & -0.325 & -0.333 \\
     & $z$ & 0.224 & 0.225 & 0.225 & 0.229 & 0.230 & 0.223 & 0.231 & 0.227 & 0.227 & 0.236 & 0.234 & 0.231 \\
%\hline
F(4) & $y$ & -0.077 & -0.079 & -0.078 & -0.078 & -0.080 & -0.080 & -0.080 & -0.076 & -0.079 & -0.082 & -0.081 & -0.081 \\
     & $z$ & 0.011 & 0.018 & 0.016 & -0.001 & -0.001 & 0.006 & -0.001 & 0.007 & 0.011 & -0.002 & 0.000 & 0.007 \\
\hline
Ba   & $d$ [\AA]& 0.256 & 0.276 & --- & 0.203 & 0.202 & --- & 0.190 & 0.238 & --- & 0.152 & 0.181 & --- \\ 
F(1) & $d$ [\AA]& 0.751 & 0.775 & --- & 0.607 & 0.695 & --- & 0.467 & 0.501 & --- & 0.329 & 0.409 & --- \\
\end{tabular}
\end{ruledtabular}
\end{table*}
\endgroup

Table~\ref{tab:struc} shows the structural parameters for all systems obtained
here by using both LSDA and LSDA+$U$ with $U_\text{eff}$ = 4\,eV, together
with experimental data. The overall agreement between calculated and
experimental values is very good. The use of the LSDA leads to a typical
underestimation of the lattice parameters between 1--4\,\% whereas the
internal structural parameters are very close to the experimental values. The
LSDA+$U$ method leads to a slightly larger equilibrium volume compared to the
pure LSDA and therefore improves the agreement with experiment. This is very
similar to what has been reported previously for several magnetic
oxides.\cite{Neaton_et_al:2005, Bandyopadhyay_et_al:2004, Bengone_et_al:2002}

The calculated structural parameters shown in Table~\ref{tab:struc} are
calculated for the experimentally observed antiferromagnetic order (see
Sec.~\ref{sec:mag}). If a different magnetic structure is imposed during the
relaxation (results not shown here), we notice small, but distinct, structural
changes. For example, relaxation of BaNiF$_4$ in a ferromagnetic configuration
leads to a 0.8\,\% increase in the lattice parameter $b$ and up to 7\,\%
change in the Wyckoff positions of those fluorine ions that mediate the
magnetic superexchange interactions in the $M$-F-$M$ bonds. This indicates a
certain degree of spin-lattice coupling in these systems which can give rise
to phenomena such as spin-phonon coupling \cite{Sushkov_et_al:2005} and
magneto-capacitance.\cite{Huang_et_al:1997}

\begin{figure}[hbpt]
\centerline{\includegraphics[width=\columnwidth]{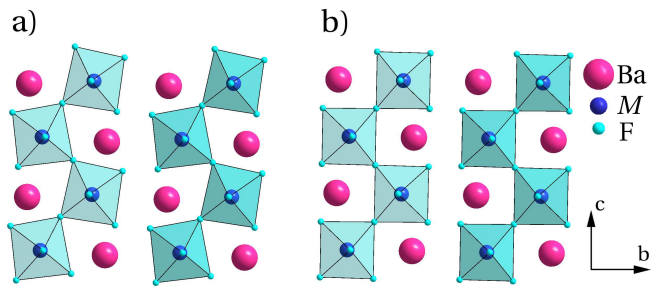}}
\caption{(Color online) a) Projection of the Ba$M$F$_4$ structure
  along the $a$ axis. The $M$ cations are octahedrally surrounded by fluorine
  anions, which form puckered sheets perpendicular to the $b$ axis, separated
  by similar sheets of Ba cations. Adjacent sheets are shifted relative to
  each other by half a lattice constant along the $a$ direction. b)
  Corresponding centrosymmetric prototype structure.}
\label{fig:struc}
\end{figure}

We also performed structural relaxations for all systems within the
corresponding centrosymmetric prototype structure with space group symmetry
$Cmcm$ (see Ref.~\onlinecite{Keve/Abrahams/Bernstein:1969} and
Fig.~\ref{fig:struc}b), which can be obtained from the ground state structure
by imposing an additional mirror symmetry perpendicular to the $c$ axis. It
has been suggested that the $M$-F(1)-$M$ distance in this centrosymmetric
$Cmcm$ structure is too small to accommodate two $M$-F bonds, causing the
structural instability in the Ba$M$F$_4$
systems.\cite{Keve/Abrahams/Bernstein:1969} Here, F(1) is the fluorine anion
connecting neighboring octahedra along the $c$ direction. The $M$-F(1)-$M$
segment thus ``bends'' outward in order to increase the $M$-F(1) bond length,
leading to the observed collective distortion of the octahedral network. This
distortion can be understood as collective alternate rotations of the fluorine
octahedra around their respective centers with the rotation axis parallel to
the $a$ direction, accompanied by displacements of the Ba cations parallel to
the polar $c$ axis. Table~\ref{tab:struc} shows the magnitude of the
displacements $d$ of both the Ba and F(1) ions leading from the
centrosymmetric reference structure to the polar ground state structure. From
these displacements it can be seen that the structural distortion increases
over the series from $M$ = Ni to $M$ = Mn, which is also consistent with the
experimental observations, and has been explained by a decrease in the ratio
between the $M$-$M$ distance and the minimum $M$-$F$ bond length when $M$ is
changed from Ni to Mn.\cite{Keve/Abrahams/Bernstein:1969}

\subsection{Electronic structure}
\label{sec:ldau}

\begin{figure}
\includegraphics*[width=0.8\columnwidth]{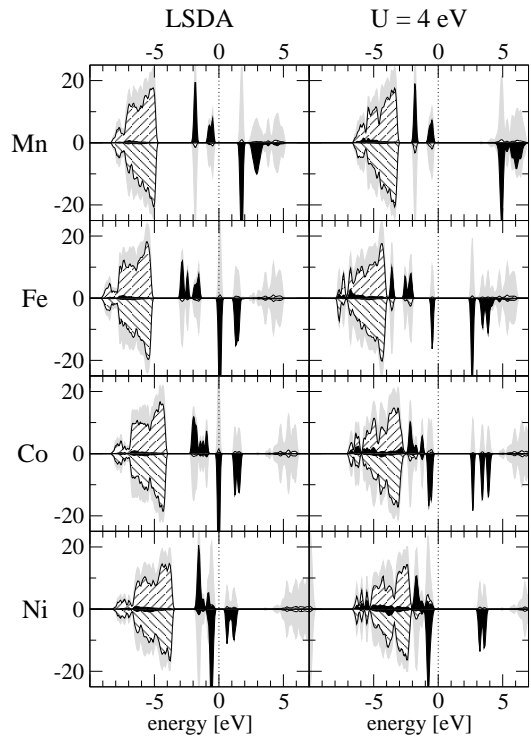}
\caption{Total densities of states (gray shaded), partial fluorine $p$
  (shaded with thin diagonal lines) and transition metal $d$ (black shaded)
  states (in states/eV), calculated within the LSDA (left panel) and by using
  $U_\text{eff}$ = 4\,eV (right panel) for all Ba$M$F$_4$ systems ($M$ = Mn,
  Fe, Co, Ni from top to bottom). Minority spin states are shown with negative
  sign. Zero energy corresponds to the Fermi level (metallic systems) or the
  highest occupied state (insulating systems).}
\label{fig:DOS}
\end{figure}

The calculated total densities of states as well as the partial densities of
the fluorine $p$ and transition metal $d$ states for all systems are shown in
Figure~\ref{fig:DOS}. For $M$ = Mn and Ni, even the use of the LSDA leads to
an insulating solution. In the case of the Ni system the gap is small and is
due to the crystal field splitting of the localized $d$ states, whereas in the
Mn system the gap is larger and is produced mainly by the strong exchange
splitting between the two spin channels. The Fe and Co systems are metallic in
LSDA, as expected for the $d^6$ and $d^7$ electron configurations of the
Fe$^{2+}$ and Co$^{2+}$ ions within the predominantly cubic crystal field. The
very small band-width of the transition metal $d$ states indicates an
instability towards the formation of a Mott-Hubbard gap, and indeed the use of
the LSDA+$U$ method with $U_\text{eff}$ = 4\,eV leads to the formation of a
large gap of about 2--3\,eV for $M$ = Fe and Co, and to a substantial increase
in the width of the gap for the Mn and Ni systems. In all cases the gap is
between occupied and unoccupied $d$ states of the transition metal ion.

These results indicate that the use of the LSDA is inadequate for the
Ba$M$F$_4$ systems, whereas the LSDA+$U$ method with an appropriate $U$ value
leads to a good description of the electronic structure. In this work, except
where otherwise noted, we use $U_\text{eff}$=4\,eV, which is a typical value
for transition metal cations in oxide
materials.\cite{Solovyev/Hamada/Terakura:1996, Pickett/Erwin/Ethridge:1998,
  Cococcioni/Gironcoli:2005} Since the overall features of the transition
metal $d$ states in the present fluorides are very similar to the oxide case,
we expect that the same value of $U$ is also appropriate for these systems. We
point out that a variation of $U$ within reasonable limits alters the
electronic structure only qualitatively. Nevertheless, for quantities that are
expected to depend critically on $U$, we always consider the explicit $U$
dependence.

It can be seen from Fig.~\ref{fig:DOS} that all transition metal cations are
in a high-spin configuration, where the local majority spin states are fully
occupied, and the minority spin states are filled with 0, 1, 2, 3 electrons
for $M$ = Mn, Fe, Co, Ni, respectively. In the case of the LSDA, the
transition metal $d$ states are energetically well separated from the fluorine
$p$ states, leading to only negligible hybridization between the two sets of
states. The use of LSDA+$U$ lowers the energy of the filled transition metal
$d$ states, leading to energetic overlap of these states with the fluorine $p$
levels and a certain degree of hybridization, which is most notable for $M$ =
Ni. This increase in hybridization with increasing $U$ could be an artifact of
the LSDA+$U$ method, which only corrects the transition metal $d$ states while
leaving the fluorine $p$ states unchanged.

The densities of states for the centrosymmetric $Cmcm$ structures (not shown
here) are indistinguishable from those calculated for the ground state
structures (both in LSDA and LSDA+$U$). This indicates that the structural
distortions in these systems do not lead to a significant change in covalent
bonding, in contrast to the case of the ferroelectric perovskites, where the
structural distortions lead to strong rehybridization between filled anion $p$
and empty cation states.\cite{Cohen:1992,Filippetti/Hill:2002} We will further
analyze the ferroelectric instability of the Ba$M$F$_4$ systems in
Sec.~\ref{sec:ferro}.

\subsection{Magnetic properties}
\label{sec:mag}

\begin{figure}
\includegraphics[width=0.5\columnwidth]{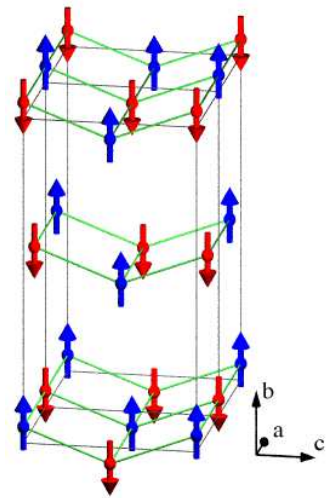}
\caption{(Color online) Magnetic structure of Ba$M$F$_4$, $M$ = Mn,
  Fe, Ni (only magnetic ions are shown). For $M$ = Co (phase A) the magnetic
  moments are oriented parallel to the $c$ axis, but the relative orientations
  of the moments are the same. Gray lines outline the conventional
  orthorhombic unit cell; the puckered rectangular grids are shown in green.}
\label{fig:mag}
\end{figure}

The experimentally observed magnetic structure is shown in
Fig.~\ref{fig:mag}.\cite{Cox_et_al:1970} The magnetic $M$ cations form
quasi-two-dimensional puckered rectangular grids ``parallel'' to the $a$-$c$
planes. Within each rectangular grid the magnetic moments of nearest neighbors
are aligned antiparallel to each other. The coupling between adjacent grids
effectively cancels, so that the magnetic order along the $b$ direction is
determined by the weak coupling between the next nearest neighbor planes that
are $\sim$14~\AA\ apart.  This can lead to two different magnetic phases, in
which the coupling along the $b$ direction is either parallel (phase B) or
antiparallel (phase A). Both phases have been observed for
BaCoF$_4$,\cite{Eibschuetz_et_al:1972} and since the energy difference between
these two phases is very small, it is generally assumed that extrinsic effects
such as defects etc. can lead to the preferred appearance of one or the other
phase. In our calculations we only consider phase A where the coupling between
second nearest neighbor planes along the $b$ direction is antiparallel. In all
systems except BaCoF$_4$ the magnetic moments are oriented parallel to the $b$
axis, whereas in BaCoF$_4$ the magnetic moments are oriented parallel to the
$c$ axis (in both phase A and phase B).\cite{Eibschuetz_et_al:1972}

The magnetic structure of BaMnF$_4$ exhibits two distinct variations of the
structure described in the previous paragraph and shown in Fig.~\ref{fig:mag}:
first, the antiferromagnetic axis is rotated by 9\,$^\circ$ from the $b$
towards the $a$ direction,\cite{Cox_et_al:1979} and in addition all magnetic
moments are slightly canted towards the $c$ direction by about 0.1\,$^\circ$,
resulting in a very small magnetization.\cite{Venturini/Morgenthaler:1974}

\begin{figure}
\centerline{\includegraphics*[width=0.7\columnwidth]{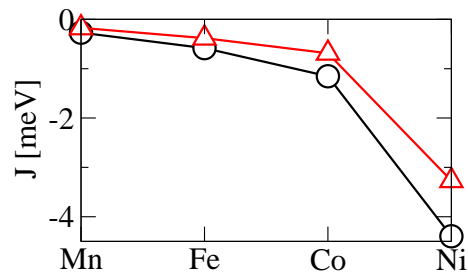}}
\caption{Heisenberg nearest neighbor exchange coupling constants $J_a$
  (circles) and $J_c$ (triangles).}
\label{fig:coupling}
\end{figure}

We do not attempt to reconstruct the full magnetic structure of all Ba$M$F$_4$
systems from first principles, but to check if the LSDA+$U$ treatment of the
electron-electron interaction leads to a correct description of the magnetic
properties of the multiferroic barium fluorides, we determine the nearest
neighbor magnetic coupling constants for all systems. To do this, we first
calculate the energy differences corresponding to different magnetic
configurations of the four spins in the unit cell. We then map the calculated
energies on a simple Heisenberg model, where we write the magnetic interaction
as $E_{ij} = -2J_{ij} s_i \cdot s_j$ ($s_i$ is the spin of cation $i$;
$J_{ij}$ is the coupling constant between ions $i$ and $j$) and consider only
nearest neighbor interactions. Within our sign convention a negative $J_{ij}$
corresponds to antiferromagnetic coupling.

Fig.~\ref{fig:coupling} shows the calculated Heisenberg coupling constants
$J_a$ and $J_c$ corresponding to pairs of nearest neighbor spins along the $a$
and $c$ direction, respectively, for $M$ = Mn, Fe, Co, and Ni.  All nearest
neighbor couplings are antiferromagnetic, in agreement with the experimentally
observed magnetic structure. The coupling becomes stronger from $M$ = Mn to
Ni, which is a consequence of the successive filling of the $t_{2g}$ states
(see Ref.~\onlinecite{Anderson:1963}). In addition, the stronger hybridization
between the transition metal $d$ and fluorine $p$ states in the case of the Ni
system further increases the strength of the superexchange interaction,
leading to particularly strong antiferromagnetic nearest neighbor coupling in
BaNiF$_4$.

\begin{figure}
\centerline{\includegraphics*[width=0.7\columnwidth]{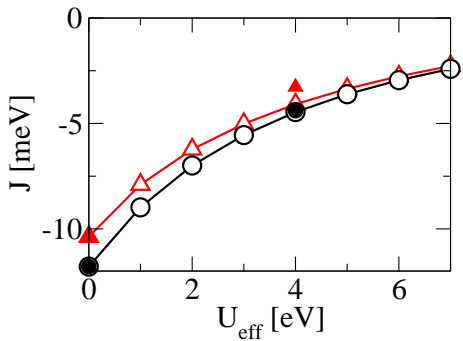}}
\caption{Heisenberg coupling constants for BaNiF$_4$ as a function of
  $U_\text{eff}$ ($J_a$: circles, $J_c$: triangles). Open symbols correspond
  to calculations done with structural parameters obtained within LSDA. Filled
  symbols correspond to fully relaxed calculations.}
\label{fig:coup_U}
\end{figure}

For BaMnF$_4$ we obtain the exchange coupling constants $J_a$ = $-$0.270\,meV
and $J_c$ = $-$0.173\,meV (using $U_\text{eff}$ = 4\,eV). This compares
extremely well with experimental values that have been extracted from the
measured spin-wave dispersion: $J_a$ = $-$0.282\,meV and $J_c$ = $-$0.197\,meV
(Ref.~\onlinecite{Shapiro_et_al:1976}). This very good agreement might be to
some extent fortuitous, but can also be regarded as strong indication that the
value of $U_\text{eff}$ indeed provides a good description of the electronic
structure, since the magnetic superexchange coupling depends very strongly on
$U$ (see Fig.~\ref{fig:coup_U} and the following paragraph).

From the theory of superexchange it follows that the corresponding coupling
constant is proportional to 1/$U$.\cite{Anderson:1963} One therefore expects a
rather strong influence of the Hubbard parameter on the strength of the
magnetic coupling, if this coupling is mediated mainly by the superexchange
interaction. Indeed, strong $U$ dependence of the magnetic coupling constants
has been found for various oxides.\cite{Baettig/Ederer/Spaldin:2005,
  Novak/Rusz:2005, Solovyev:2002} Fig.~\ref{fig:coup_U} shows the variation of
$J_a$ and $J_c$ in BaNiF$_4$ with the Hubbard parameter $U_\text{eff}$ used in
the LSDA+$U$ treatment of the electron-electron interaction. To see the pure
electronic effect on the coupling constant, the structural parameters are kept
fixed to their values obtained by the LSDA relaxation. It is apparent that the
coupling strength significantly decreases with increasing $U_\text{eff}$,
although the variation does not show an exact 1/$U$ dependence. This is due to
the fact that a variation of $U$ in the bandstructure calculation also changes
the overlap of the wave-functions and therefore the transfer integrals, which
are considered as constant in the theory of superexchange.

The values corresponding to the relaxed structural parameters for
$U_\text{eff}$ = 4\,eV are also shown in Fig.~\ref{fig:coup_U}. It can be seen
that there is only a negligible change of $J_a$, whereas $J_c$ is reduced by
20\,\% due to the slightly different structure. This is due to the fact that
the bond lengths and angles for the Ni-F(4)-Ni bond, which determines $J_a$,
are nearly identical in LSDA and LSDA+$U$, whereas the Ni-F(1)-Ni bond angle,
which determines $J_c$, is reduced from 160.5\,$^\circ$ (LSDA) to
155.8\,$^\circ$ ($U_\text{eff}$ = 4\,eV). The Ni-F(1) distance increases
slightly from 1.94\,\AA\ (LSDA) to 1.96\,\AA\ ($U_\text{eff}$ = 4\,eV).

In addition, if spin-orbit coupling is taken into account, there are small
deviations from the collinear antiferromagnetic
structure,\cite{Ederer/Spaldin:2006} which are due to the antisymmetric
exchange or Dzyaloshinskii-Moriya interaction.\cite{Moriya:1963} This small
canting does not influence the structural and ferroelectric properties studied
in the present paper, and we therefore neglect all effects due to spin-orbit
coupling here.

\subsection{Ferroelectric properties}
\label{sec:ferro}

\begin{figure}
\includegraphics*[width=0.7\columnwidth]{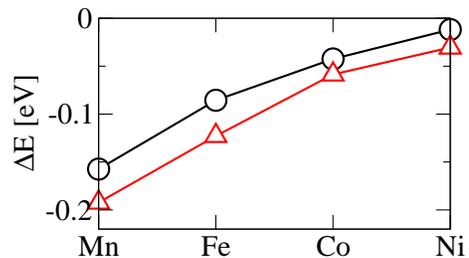}
\caption{Energy differences $\Delta E$ per formula unit between the
  ferroelectric and the centrosymmetric prototype structures, calculated
  within the LSDA (circles) and by using $U_\text{eff}$=4\,eV (triangle).}
\label{fig:ediff}
\end{figure}

Fig.~\ref{fig:ediff} shows the energy differences per formula unit between the
polar ground-state $Cmc2_1$ structures and the corresponding centrosymmetric
$Cmcm$ structures, calculated within the LSDA and by using LSDA+$U$ with
$U_\text{eff}$ = 4\,eV. The latter results in slightly larger energy
differences, but the overall trend is similar in the two cases. The energy
difference is largest for $M$ = Mn and smallest for $M$ = Ni, consistent with
the trend in the magnitude of the structural distortions (see
Sec.~\ref{sec:struc}). The fact that no ferroelectric switching could be
observed for Ba$M$F$_4$ with $M$ = Mn and Fe \cite{Eibschuetz_et_al:1969} has
been explained by suggesting a rather high energy for the intermediate
centrosymmetric state in these two systems, which results from the different
equilibrium bond lengths compared to the systems with $M$ = Co, Ni, Zn, and
Mg.\cite{Keve/Abrahams/Bernstein:1969} This energy barrier is exactly the
energy difference shown in Fig.~\ref{fig:ediff}. The values for $M$ = Ni and
$M$ = Mn are comparable with the corresponding energy differences between
ferroelectric and centrosymmetric prototype structures in the perovskite
ferroelectrics BaTiO$_3$ ($\approx$ 18\,meV/f.u.) and PbTiO$_3$ ($\approx$
200\,meV/f.u.), respectively.\cite{Cohen:1992} Although it is not likely that
ferroelectric switching actually takes place through the centrosymmetric
reference structure, this energy difference can be viewed as an upper limit
for the switching barrier. It is apparent that a classification of the Co and
Ni systems as ferroelectric and the Mn and Fe systems as merely pyroelectric
is probably not very instructive, since there is only a gradual quantitative
difference between these systems. It seems feasible that ferroelectric
switching could be achieved at least for the Fe system if highly resistive
samples were available that could sustain higher electric fields.

\begin{table}
\caption{Spontaneous polarizations (in $\mu$C/cm$^2$) calculated with
  an ionic model using formal charges and by using the Berry-phase
  approach ($U_\text{eff}$ = 4~eV), together with available
  experimental data.}
\label{tab:pol}
\begin{ruledtabular}
\begin{tabular}{cccc}
 & ionic model & Berry phase & Exp. [\onlinecite{Eibschuetz_et_al:1969}] \\
\hline
Mn & 13.68 & 13.60 & --- \\
Fe & 10.34 & 10.88 & --- \\
Co & 8.57 & 9.02 & 8.0 \\
Ni & 6.34 & 6.79 & 6.7 
\end{tabular}
\end{ruledtabular}
\end{table}

From the displacements of the ions between the centrosymmetric prototype and
the polar ground state structure it is possible to calculate an estimate of
the spontaneous polarization using the formal charges of the ions (2+ for the
cations and 1-- for the fluorine anions). Table~\ref{tab:pol} lists the
polarization values calculated from such an ionic model together with the
corresponding values obtained by \emph{ab initio} calculation using the
Berry-phase theory of electric polarization.\cite{King-Smith/Vanderbilt:1993,
  Vanderbilt/King-Smith:1993, Resta:1994} The good agreement between the
Berry-phase result and the ionic model indicates that the Born effective
charges \cite{Ghosez/Michenaud/Gonze:1998} are very close to their formal
values for all systems. This is consistent with the lack of significant
changes in the electronic densities of states between the centrosymmetric and
distorted structures as described in Sec.~\ref{sec:ldau}. Thus, no significant
charge transfer occurs as a result of the polar distortion. This again
indicates that the ferroelectricity in these systems is of different origin
than in the prototypical perovskite ferroelectrics where charge transfer
between the transition metal $d$ and the anion $p$ states is crucial for
stabilizing the ferroelectric state.\cite{Cohen:1992}

\begin{table}
\caption{TO phonon frequencies $\omega$ of $A_g$ and $B_{1u}$ symmetry
  for BaNiF$_4$. $x$ gives the contribution of the corresponding mode
  to the ground state distortion.}
\label{tab:phonon}
\begin{ruledtabular}
\begin{tabular}{llcccccc}
$A_g$ & $\omega$ [cm$^{-1}$] & 96 & 168 & 256 & 300 & 502 & \\
      &      $x$ [\%]        &  7 &  3  &  1  &  0  &  0  & \\
\hline
$B_{1u}$ & $\omega$ [cm$^{-1}$] & $i\cdot$58 & 151 & 212 & 256 & 397 & 541 \\
         & $x$ [\%]             &  87        & 1   &  0  &  0  &  0  & 0   \\
\end{tabular}
\end{ruledtabular}
\end{table}

To obtain further insight into the ferroelectric instability of these systems,
we calculate the TO phonon frequencies of the centrosymmetric structure for
BaNiF$_4$. Since we are only interested in the phonon modes that are
compatible with the observed ground state symmetry (only these modes can give
rise to the corresponding structural distortions), we transform the dynamical
matrix to block-diagonal form using symmetry-adapted modes, and diagonalize
only the blocks corresponding to the irreducible representations $A_g$ and
$B_{1u}$ of space group $Cmcm$. The nonpolar $A_g$ phonons do not change the
symmetry of the system, whereas the infrared active $B_{1u}$ modes reduce the
symmetry to the ground state $Cmc2_1$ space group. There are 5 $A_g$ and 7
$B_{1u}$ modes. The calculated phonon frequencies are listed in
Table~\ref{tab:phonon}. One of the 7 $B_{1u}$ modes corresponds to an acoustic
mode with zero frequency at the $\Gamma$ point and is not included in
Table~\ref{tab:phonon}.

We find that there is one unstable $B_{1u}$ mode with $\omega =
i\cdot$58\,cm$^{-1}$. If we decompose the structural distortion leading from
the centrosymmetric reference structure to the relaxed ground state structure
into contributions from the different phonon modes, we see that the unstable
$B_{1u}$ mode is responsible for about 87\,\% of the final distortion. This
indicates that the ferroelectricity in the barium fluorides originates from
the softening of a single polar phonon mode which ``freezes in'' to form the
ferroelectric ground state of the system. In contrast to the well-known case
of perovskite ferroelectrics such as BaTiO$_3$ or PbTiO$_3$ no charge transfer
between anions and cations occurs as a result of the displacements, which
would lead to anomalous values of the Born effective charges, but instead the
structural distortion in the Ba$M$F$_4$ systems is driven purely by size
effects, which, together with the special geometric connectivity realized in
these compounds, leads to an inversion symmetry breaking and the appearance of
a spontaneous electric polarization.

\section{Discussion and Summary}

It is well known that size-effects can drive structural distortions. For
example in the perovskite system the relative ionic radii of the $A$ and $B$
cations determine the stability towards tilts of the anion
octahedra.\cite{Goldschmidt:1926}

The idea that ionic polarizability is, in general, not necessary for achieving
ferroelectricity was already formulated in
Ref.~\onlinecite{Edwardson_et_al:1989}, when a ferroelectric ground state was
predicted for perovskite-like NaCaF$_3$ on the basis of atomistic simulations
using Gordon-Kim pair-potentials.\cite{Gordon/Kim:1972} Since the possibility
of covalent bond formation is excluded within this method, it became clear
that a different mechanism is driving the ferroelectric distortion in
NaCaF$_3$. Although the underlying mechanism was not investigated any further,
it was assumed that ionic polarizability would significantly enhance the
ferroelectricity in NaCaF$_3$. The present study shows that this does not
necessarily have to be the case. Ferroelectricity was also predicted recently
for a variety of perovskite-like fluorides, such as LiMgF$_3$, LiNiF$_3$, and
NaCdF$_3$.\cite{Claeyssens_et_al:2003,Duan_et_al:2004} It is reasonable to
assume that the ferroelectricity in all these fluoride systems is driven
purely by size effects and that no charge transfer between anions and cations
occurs.

It was pointed out recently that such a mechanism for ferroelectricity is
compatible with the simultaneous occurrence of magnetic
ordering.\cite{vanAken_et_al:2004} Using first principles electronic structure
calculations, the hexagonal multiferroic YMnO$_3$ was identified as an example
for such a scenario. In this material no charge transfer occurs as a result of
the structural distortion, and the dynamical charges are very close to their
formal values.\cite{vanAken_et_al:2004} This suggests that in YMnO$_3$,
similar to the Ba$M$F$_4$ systems discussed in the present paper, ionic
polarizability does not have any effect on the ferroelectric properties. A
subsequent analysis of the phonon modes of the corresponding centrosymmetric
reference structure suggested that YMnO$_3$ is in fact an \emph{improper
  ferroelectric}, where the ferroelectricity is due to the symmetry-allowed
coupling of a stable polar zone-center phonon mode to an unstable
zone-boundary mode.\cite{Fennie/Rabe_YMO:2005}

The present study shows that such ``geometric ferroelectricity'' is not
restricted to improper ferroelectrics, but that size effects can also directly
lead to instable polar zone center phonon modes. Thus, BaNiF$_4$, and with it
the whole class of Ba$M$F$_4$ multiferroics, represent the first confirmed
example of a \emph{proper} geometric ferroelectric.

In summary, we have presented an \emph{ab initio} study of the structural,
electronic, and magnetic properties of the multiferroic barium fluorides
Ba$M$F$_4$ with $M$ = Mn, Fe, Co, and Ni. The ferroelectricity in these
systems is due to a single unstable polar zone-center phonon. The instability
is triggered solely by size effects, and no charge transfer occurs as a result
of the structural distortion. In addition, we have shown that the LSDA+$U$
method results in a good description of the electronic structure of these
systems. Our work represents the first \emph{ab initio} demonstration of a
proper geometric ferroelectric and the first such study of a non-oxide
multiferroic.

\begin{acknowledgments}
  This work was supported by the NSF's \emph{Chemical Bonding Centers}
  program, Grant No. CHE-0434567 and made use of the central facilities
  provided by the NSF-MRSEC Award No. DMR05-20415. The authors thank Craig
  J.~Fennie and J.~F.~Scott for valuable discussions.
\end{acknowledgments}

\bibliography{references.bib}

\end{document}